\documentclass[doublespacing]{elsart}

\usepackage{amssymb}
\usepackage{amsmath}
\usepackage{graphicx}

\begin{document}

\begin{frontmatter}



\title{Joint effects of nutrients and contaminants on the
dynamics of a food chain in marine ecosystems}


\author[ifisc]{Flora S. Bacelar,}
\author[jrc]{Sibylle Dueri,}
\author[ifisc]{Emilio Hern\'{a}ndez-Garc\'{\i}a\corauthref{corr},}
\corauth[corr]{Corresponding author. E-mail: emilio@ifisc.uib.es}
\author[jrc]{Jos\'{e}-Manuel Zald\'{\i}var}

\address[ifisc]{IFISC, Instituto de F{\'\i}sica Interdisciplinar y
Sistemas Complejos (CSIC-UIB), Campus Universitat de les Illes
Balears, E-07122 Palma de Mallorca, Spain}
\address[jrc]{European Commission, Joint Research Centre, Via E. Fermi 2749, TP 272
I-21027 Ispra (VA), Italy}

\begin{abstract}
We analyze the joint effect of contaminants and nutrient loading
on population dynamics of marine food chains by means of
bifurcation analysis. Contaminant toxicity is assumed to alter
mortality of some species with a sigmoidal dose-response
relationship. A generic effect of pollutants is to delay
transitions to complex dynamical states towards higher nutrient
load values, but more counterintuitive consequences arising from
indirect effects are described. In particular, the top predator
seems to be the species more affected by pollutants, even when
contaminant is toxic only to lower trophic levels.
\end{abstract}


\end{frontmatter}

\section{Introduction}
\label{}

Marine waters and in particular coastal waters are increasingly
exposed to anthropogenic pressures represented not only by the
growing input of nutrients and contaminants related to urban,
agricultural and industrial activities, but also by the
exploitation of coastal areas for aquaculture, fishing and
tourism. Since the resources of the coastal zone are exploited by
different stakeholders, it is essential to improve our knowledge
on the ecosystem's vulnerability to stressors and protect those
areas through a sensible management.

The interaction of pollutants and nutrients on aquatic ecosystems
is difficult to evaluate, since many direct and indirect effects
have to be considered. Contaminants can have instantaneous
effects, such as massive killings after an accidental contaminant
release. Other toxic effects, such as genotoxicity and
reproductive failure are less evident and they act on a longer
time-scale; however, they represent an important risk for the
ecosystem. Furthermore, if the contaminant is lipophilic,
bioaccumulation should be considered. On the other hand, an
increase of the nutrient load can have the direct effect of
raising the primary production at the bottom of the food chain and
consequently increasing the concentration of the organic matter in
the system. But a higher concentration of organic matter can
affect the bioavailability of the contaminants and therefore the
fate of pollutants in the aquatic environment and their effects on
the impacted ecosystem \cite{Gunnarsson1995}.

Thus, contaminants affect aquatic ecosystems through direct and
indirect effects \cite{Fleeger2003}, from acute and/or chronic toxicity on
sensitive species to disruption in the food web structure. Some
species might be more sensitive than others to a particular chemical,
but since the different populations are linked to each other by
competition and predation, species which are not directly stressed
may respond indirectly \cite{Fleeger2003}. Within a food web, community-level
relations arise from unobservable indirect pathways. These
relations may give rise to indirectly mediated relations,
mutualism and competition \cite{Fath2007}. In some cases environmental
perturbations alter substantially the dynamics or the structure of
coastal ecosystems and the effect may produce the occurrence of a
trophic cascade and eventually the extinction of some species \cite{Jackson2001}.
A better understanding of the relative importance of top-down
(e.g. overfishing) versus bottom-up (e.g. increased nutrient input
causing eutrophication) controls is essential and can only be
achieved through modelling \cite{Daskalov2002}.

Sudden regime shifts and ecosystem collapses are likely to occur
in stressed ecosystems. Catastrophic regime shifts have been
related to alternative stable states which can be linked to a
critical threshold in such a way that a gradual increase of one
driver has little influence until a certain value is reached at
which a large shift occurs that is difficult to reverse \cite{Scheffer2003,Scheffer2001}. The
shape of ecotoxicological dose-response curves \cite{Suter1993}, showing a
sharp increase in the effect of toxic substances above a critical
value, facilitates the occurrence of regime shifts under pollutant
pressure.

In this study we consider the combined effects of contaminant
substances and nutrient load in the framework of a simple
tritrophic food chain model. We restrict our study to
contaminants, such as s-triazines, which affect the mortality in
particular trophic levels, but which do not bioaccumulate in time
nor along the food chain. When studying the dynamics of simple
food chain and food web models, it is also important to bear in
mind that the response might depend on the complexity of the
represented system. Chaotic dynamics, for example, seems to be
more frequent in simple ecosystem models or in models with a high
number of trophic levels \cite{Fussmann2002}. Thus, we will focus
only on the first qualitative changes of behaviour occurring when
increasing nutrients from low values, and how this is changed by
pollutants, and not on the complex sequences of chaotic states
which may occur at high nutrient availability, whose details are
more affected by the trophic structure of the model.

Since we do not include any microbial recycling loop, sediment or
oxygen dynamics, or shading effects, complex eutrophication
behaviour typical of coastal ecosystems \cite{Zaldivar2003}, e.g.
anoxic crises, alteration of nutrient cycling, macroalgal blooms,
etc, will not occur in our model. We rather concentrate on the
simplest scenarios occurring during enrichment and its
modification by contaminants, discussing particularly the indirect
effects which lead to counterintuitive behaviour.

\section{Model formulation}

We consider \cite{Boer1998,Boer2001,Gwaltney2007} Canale's
chemostat model (CC), which is an extension of the tri-trophic
food-chain Rosenzweig-MacArtur model (RMA) that has been
extensively studied in theoretical ecology
\cite{Hastings1991,Klebanoff1994,Abrams1994,McCann1994,Kuznetsov1996,DeFeo1997,Gragnani1998,Gwaltney2004}.
This model was previously used to analyze the dynamics of a food
chain consisting of bacteria living on glucose, ciliates and
carnivorous ciliates \cite{Boer1998,Boer2001}, but can be adapted
to represent an aquatic food chain with a constant nutrient input.
The CC model is similar to the RMA model, but there is an
additional equation representing the input of nutrient, and it
considers the losses due to the flushing rate:
\begin{eqnarray}
  \dot{N} &=& D(I-N)-P\frac{a_{1}N}{b_{1}+N},\label{1-CC}\\
  \dot{P} &=& P\left[e_{1}\frac{a_{1}N}{b_{1}+N}-\frac{a_{2}Z}{b_{2}+P}-d_{1}-f_{1}D \right],\label{2-CC}\\
  \dot{Z} &=& Z\left[e_{2}\frac{a_{2}P}{b_{2}+P}-\frac{a_{3}F}{b_{3}+Z}-d_{2}-f_{2}D\right],\label{3-CC}\\
  \dot{F}
&=&F\left[e_{3}\frac{a_{3}Z}{b_{3}+Z}-d_{3}-f_{3}D\right]\label{4-CC}.
\end{eqnarray}

\begin{figure}[p]
\begin{center}
    \includegraphics[width=0.8\columnwidth,angle=0]{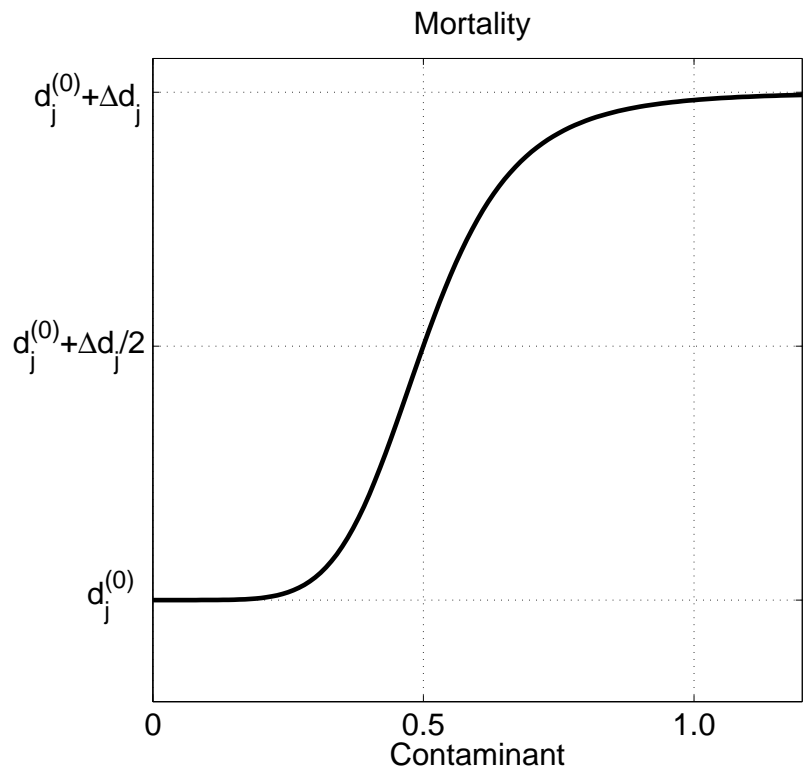} \quad
       \caption{Sigmoidal response of mortality to the concentration of
       the toxic contaminant, according to Eq. (6).}
    \label{fig1-contaminant}
\end{center}
\end{figure}

In this study the variables $N$, $P$, $Z$, $F$ represent the
nitrogen concentration in the different compartments of the system
(nutrient, phytoplankton, zooplankton and fish, which will be also
denoted with the alternative names of nutrient, prey, predator,
and top-predator, respectively) expressed in units of $mg N/l$.
Our default parameters (see Table \ref{parameter-CC}) are from
\cite{Jorgensen1979}, as used in the aquatic food chain model
presented in \cite{Mosekilde1998}. Apart from unessential scaling
differences, most of the parameters are of the same order as in
\cite{Gragnani1998,Gwaltney2007}, except that we use larger
mortality values and, accordingly, smaller flushing rates to avoid
complete extinction of the ecosystem. Our base mortalities and the
rest of parameters are consistent with the ones used in the
pelagic ecosystem model in \cite{Lima2002} which, as discussed in
that reference, are adequate for the oligotrophic subtropical
ocean. $I$ is the nutrient load or nutrient input into the system.
$D$ is a flow rate quantifying water renewal in the system, which
affects the species through the flushing rates $f_{i}$
($i=1,2,3$). $d_i$ are the specific mortalities, $b_i$ half
saturation constants for the Holling type-II predation functions,
$a_i$ are maximum predation rates, and $e_i$ efficiencies. We note
that the following condition should be satisfied by the equation
parameters:
\begin{equation}\label{food-top-pred-CC}
    e_{i}a_{i}>d_{i}+Df_{i} \ \ \ \ ( i= 1, 2, 3),
\end{equation}
since this ``avoids the case where predator and top-predator
cannot survive, even when their food is infinitely abundant" \cite{Kuznetsov2001}.
Contaminant toxicity is incorporated in our model by an increase
in mortality. We consider three different scenarios in each of
which the contaminant affects the mortality of only one of the
compartments:
\begin{equation}\label{6-dj}
    d_{j}=d_{j}^{(0)}+\Delta d_{j}\left(\frac{(C_{j})^{6}}{(C_{j})^{6}+0.5^{6}}\right)
\end{equation}
$j=1,2$, and $3$ labels the three trophic levels: prey, predator
and top-predator, $C_{j}$ is the dimensionless concentration of
the contaminant affecting the level $j$, normalized in such a way
that for $C_{j} =0.5$ the toxicity achieves half its maximum
impact on mortality, and a sigmoidal function (Fig.
\ref{fig1-contaminant}) has been used to model the mortality
increase from a baseline value, $d_{j}^{(0)}$, to the maximum
mortality, $d_{j}^{(0)}+\Delta d_{j}$, attained at large
contaminant concentrations. This represents typically the shape of
the dose-response curves found when assessing toxic effects of
chemical on biological populations \cite{Suter1993}. The values of
$d_{j}^{(0)}$ and $\Delta d_{j}$ used are written in Table
\ref{parameter-contaminant}. Other works that have studied
bifurcations due to mortality changes in the CC model
\cite{Gwaltney2007} have normally considered a linear increase.
Considering a sigmoidal response allows the identification of the
range of mortality values which are to be expected in the presence
of a given contaminant, and thus permits to focus in such range.
But once the relevant interval is identified the bifurcation
behavior can be studied as a function of the mortalities $d_i$.
This was done in Ref. \cite{Gwaltney2007} with emphasis in steady
coexistence solutions. Here, in addition to exploring a different
set of base parameters and to focusing in the mortality range implied
by the contaminant characteristics in Table
\ref{parameter-contaminant}, we also perform continuations of
cyclic solutions and find some period doubling bifurcations which
would eventually lead to chaos.

\section{Steady states}

This system presents the following set of fixed points: The
nutrient-only state (Nu):
\begin{eqnarray}\label{7-N}
  N &=& I, \notag\\
  P &=& 0, \\
  Z &=& 0,\notag\\
  F &=& 0.\notag
\end{eqnarray}

The nutrient-phytoplankton state (NP):
\begin{eqnarray}\label{8-NP}
  N &=& \frac{b_{1}(d_{1}+Df_{1})}{a_{1}e_{1}-d_{1}-Df_{1}}, \notag\\
  P &=& \frac{De_{1}\left(\frac{b_{1}(d_{1}+Df_{1})}{a_{1}e_{1}-d_{1}-Df_{1}}+
  I\right)}{d_{1}+Df_{1}},\\
  Z &=& 0, \notag\\
  F &=& 0. \notag
\end{eqnarray}

There are two solutions (NPZ) characterized by the absence of the top predator:
\begin{eqnarray}\label{9-NPZ}
  N &=&  \frac{b_{1}D+a_{1}P-DI\pm\sqrt{4b_{1}D^{2}I+(-b_{1}D-a_{1}P+DI)^{2}}}{2D},\notag\\
  P &=&  \frac{b_{2}(d_{2}+Df_{2})}{a_{2}e_{2}-d_{2}-Df_{2}},\\
  Z &=&  -\frac{(b_{1}d_{1}+b_{1}Df_{1}+d_{1}N-a_{1}e_{1}N+Df_{1}N)(b_{2}+P)}{a_{2}(b_{1}+N)}, \notag\\
  F &=& 0. \notag
\end{eqnarray}
but only the one with the positive sign of the square root is
positive definite.

Finally, there are three internal fixed points (NPZF), in which
all species occur at positive densities. From the equation for
$\dot{N}$, (\ref{1-CC}), an equation for $P$  as a function of $N$
is obtained. Introducing it into (\ref{2-CC}) together with the
expression for $Z=\bar{Z}$ which is obtained from (\ref{4-CC}), we
get the following equation for $N$:
\begin{equation}\label{10-x0}
    [A_{1}N^{3}+A_{2}N^{2}+A_{3}N+A_{4}] = 0
\end{equation}
where
\begin{eqnarray}\label{11-A}
  A_{1} &=& D(a_{1}e_{1}-d_{1}-D_{0}f_{1}), \notag \\
  A_{2} &=& -a_{1}^{2}b_{2}e_{1}-D(d_{1}+Df_{1})(2b_{1}-I)+a_{1}(b_{1}De_{1}+b_{2}(d_{1}+Df_{1})+a_{2}\bar{Z}-De_{1}I),\nonumber \\
  A_{3} &=& b_{1}(-D(d_{1}+Df_{1})(b_{1}-2I)+a_{1}(b_{2}(d_{1}+Df_{1})+a_{2}\bar{Z}-De_{1}I)), \\
  A_{4} &=& b_{1}^{2}D(d_{1}+Df_{1}I).\notag
\end{eqnarray}

The values of the remaining variables at the three internal fixed
point solutions can be written in terms of $\bar{Z}$ and of the
three values of $N=\bar{N}$ obtained from the cubic (\ref{10-x0}):
\begin{eqnarray}\label{12-NPZF}
  N &=&  \bar{N} ,\notag\\
  P &=&  D_{0}(I-\bar{N})\frac{b_{1}+\bar{N}}{a_{1}\bar{N}},\notag\\
  Z &=& \bar{Z}= \frac{b_{3}(d_{3}+Df_{3})}{a_{3}e_{3}-d_{3}-Df_{3}}, \\
 F &=& \frac{(a_{2}e_{2}P-b_{2}d_{2}-b_{2}Df_{2}-d_{2}P-Df_{2}P)(b_{3}+\bar{Z})}
 {a_{3}(b_{2}+P)}. \notag
\end{eqnarray}
It turns out that only one of the three fixed point solutions is positive for
the parameter values in Table \ref{parameter-CC}.

The above are all the biologically relevant fixed points. There
are four additional mathematical steady state solutions, but some
populations take negative values on them.

\section{Stability and bifurcation analysis}

\begin{figure}[p]
\begin{center}
    \includegraphics[width=0.8\columnwidth,angle=0]{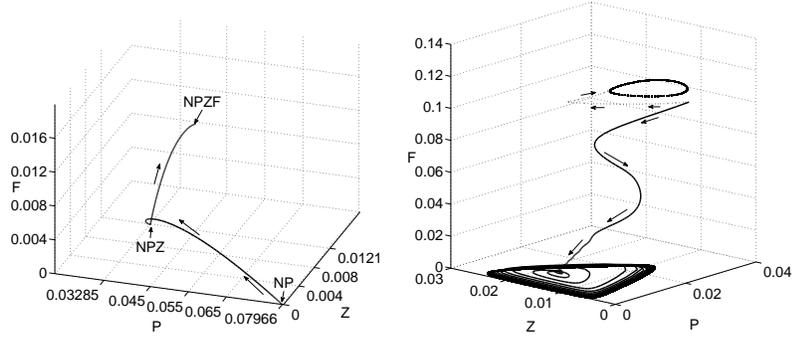}\\
    \caption{a) Projection on the PZF subspace of a trajectory which starts
close to the NP fixed point, approaches the NPZ one, and finally
is attracted by the NPZF fixed point. $I=0.4~mgN/l$, $C_1=C_3=0$,
and $C_2=0.8$. This shows the approximate location of these points
and that only the NPZF one is stable for these parameter values.
(b) Cyclic behaviour: Thick line is a trajectory leading to an
attracting limit cycle on the NPZ hyperplane for $I=0.1~mgN/l$,
$C_1=C_2=0$, and $C_3=0.8$ ; dotted line is a trajectory attracted
by the limit cycle involving all the variables for $I=0.24~mgN/l$,
$C_1=C_2=0$, and $C_3 =0.2$.}
    \label{fig2-states}
\end{center}
\end{figure}

We have analyzed the dynamics of the CC food-chain models for
several parameter values by direct numerical integration of the
model equations, and by bifurcation analysis carried on with the
software AUTO \cite{Doedel1997}. Background on the different types
of bifurcations can be found in
\cite{Strogatz2000,Gugenheimer1993}. We consider only bifurcations
of positive solutions and, as stated in the introduction, we find
but we do not describe in detail the routes to chaotic behaviour
occurring at high nutrient load. For low and intermediate nutrient
load we find that the relevant attractors are the fixed points
described above, and also two limit cycles, one involving the
variables $N$, $P$ and $Z$, lying on the $F=0$ hyperplane, and
another one in which all the species are present. These attractors
are represented in Fig.\ref{fig2-states}.

\subsection{The non-contaminant case}

\begin{figure}[p]
\begin{center}
    \includegraphics[width=.85\columnwidth,angle=0]{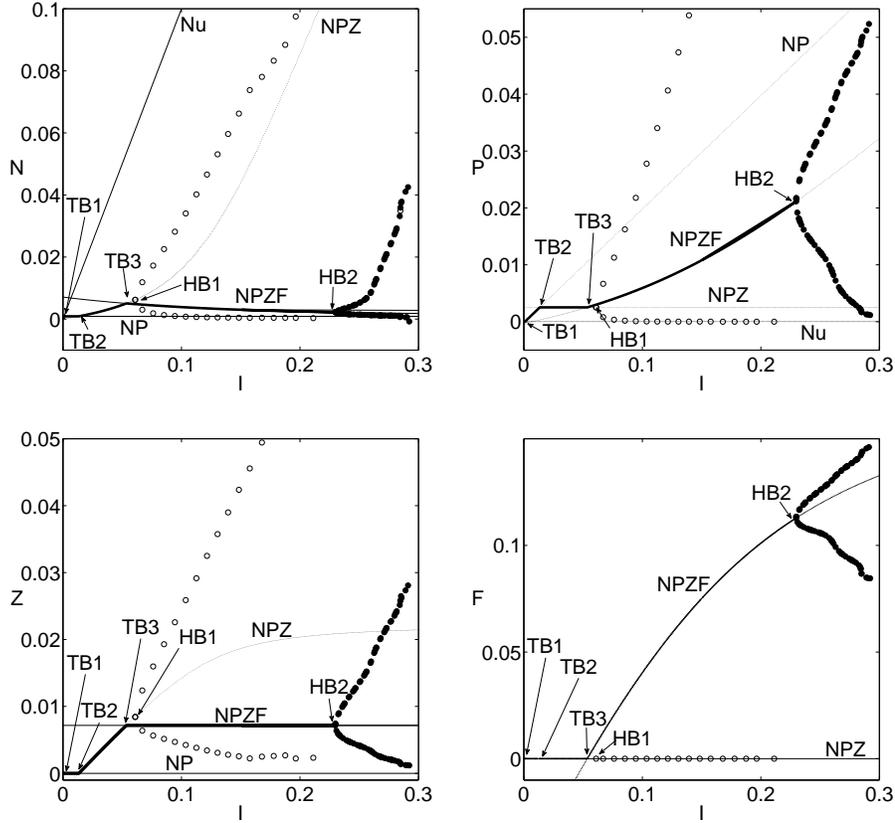}\\
   \caption{Bifurcation diagrams of the four variables as a function
   of nutrient input parameter $I$ in the absence of contaminants. Thick lines
   and full symbols denote
   stable fixed points and maxima and minima of stable cycles, respectively,
   and thin lines and open symbols, unstable ones. The name of the fixed
   points is indicated. The relevant bifurcations (described in the main text)
   occur at $I^{TB1}=0.0008909 ~mgN/l$, $I^{TB2}=0.01345 ~mgN/l$,
   $I^{TB3}= 0.05352 ~mgN/l$, $I^{HB1}= 0.06101 ~mgN/l$, and
   $I^{HB2}=0.2298 ~mgN/l$, locations which are indicated by arrows. }
   \label{fig3-nutrientes}
\end{center}
\end{figure}

First, we consider  system behaviour for the case of mortalities
at their base values, i.e. in the absence of contaminants. This
will serve as a reference for later inclusion of contaminants.
Fig.\ref{fig3-nutrientes} shows the sequence of bifurcations when increasing the
nutrient input $I$. For very low input, only nutrients are present
in the system (solution (\ref{7-N})). When $I>I^{TB1}$, with
 \begin{equation}\label{13-TB1}
  I^{TB1} = \frac{b_{1}(d_{1}+Df_{1})}{a_{1}e_{1}-d_{1}-Df_{1}}
  \ ,
 \end{equation}
phytoplankton becomes positive in a transcritical bifurcation
(which we call TB1) at which the NP state (\ref{8-NP}) becomes stable.
Since $I^{TB1}=0.0008909$ is very small, this bifurcation can not
be clearly seen in Fig. \ref{fig3-nutrientes}. From this value on, further enrichment
leads to a linear increase of phytoplankton (\ref{8-NP}), until a second
transcritical bifurcation, TB2, at which zooplankton becomes
positive and the NPZ solution (\ref{9-NPZ}) becomes stable. It happens at
 \begin{equation}\label{14-TB2}
  I^{TB2} = \frac{(d_{1}+Df_{1})(P^{NPZ}d_{1}-P^{NPZ}a_{1}e_{1}-b_{1}De_{1}+P^{NPZ}Df_{1})}
{De_{1}(d_{1}-a_{1}e_{1}+Df_{1})}
 \end{equation}
 where $P^{NPZ}$ is the expression for $P$ in the $NPZ$ solution, (\ref{9-NPZ}).
{}From this point the zooplankton starts increasing (keeping
phytoplankton concentration at a constant value) until a new
bifurcation TB3 occurs, at which the fish concentration starts to
grow from zero while zooplankton remains constant, phytoplankton
increases, and nutrients decrease (this is the positive interior
solution NPZF, Eq. (\ref{12-NPZF})). The value of $I^{TB3}$ is
given implicitly by:
\begin{equation}\label{15-TB3}
  d_{3}^{TB3} = \frac{Z^{NPZ}a_{3}e_{3}-Z^{NPZ}Df_{3}-b_{3}Df_{3}}{Z^{NPZ}+b_{3}}
 \end{equation}
 where $Z^{NPZ}$ is the expression for $Z$ in the $NPZ$ solution, (\ref{9-NPZ}).

 We note here one of the first counterintuitive indirect effects present
in the food-chain dynamics: In the NPZF solution, increase of
nutrient input leads to decrease in nutrient concentration (see
Fig. \ref{fig3-nutrientes}). The reason is the top-down control
that the higher predator begins to impose on zooplankton, leading
to positive and negative regulation on the compartments situated
one or two trophic levels below $Z$, respectively.

Shortly after becoming unstable at TB3, the fixed point NPZ
experiences a Hopf bifurcation (HB1) which leads to a limit cycle
on the NPZ hyperplane. Since the whole hyperplane has become
unstable before this bifurcation occurs, this cycle has no direct
impact on long time dynamics, although it can affect transients,
and it will become relevant when adding contaminants. The steady state
coexistence of the three species at the NPZF fixed point remains
stable until a new Hopf bifurcation HB2 occurs at which the fixed
point becomes unstable and oscillations involving the three
species and the nutrients (Fig. \ref{fig2-states}) occur. The
destabilization of steady state coexistence by the appearance of
oscillations, which could facilitate extinctions if the amplitude
of oscillation is sufficiently large, is the well known ``paradox
of enrichment", first mathematically described by Rosenzweig
\cite{Rosenzweig1971}. A good overview of the studies connected
with this issue can be found in the paper \cite{Jensen2005}. See
also \cite{Fussmann2000,Shertzer2002,Bell2002,Kirk1998}.

Gragnani et al.\cite{Gragnani1998} demonstrated that the dynamics
of Canale's model for increasing nutrient supply is qualitatively
similar to the one of the RMA model. After the stationary and
cyclic states described above, chaotic behaviour followed by a
different cyclic behaviour with higher frequency are found. Also,
the maximal average density of top-predator is attained at the
threshold between chaotic and high frequency cyclic behaviour. We do
not describe these states further but concentrate on the
modifications arising from toxic effects of contaminants on the
dynamics, for small and moderate nutrient loading.

\subsection{Contaminant toxic to phytoplankton}

\begin{figure}[p]
\begin{center}
    \includegraphics[width=.8\columnwidth,angle=0]{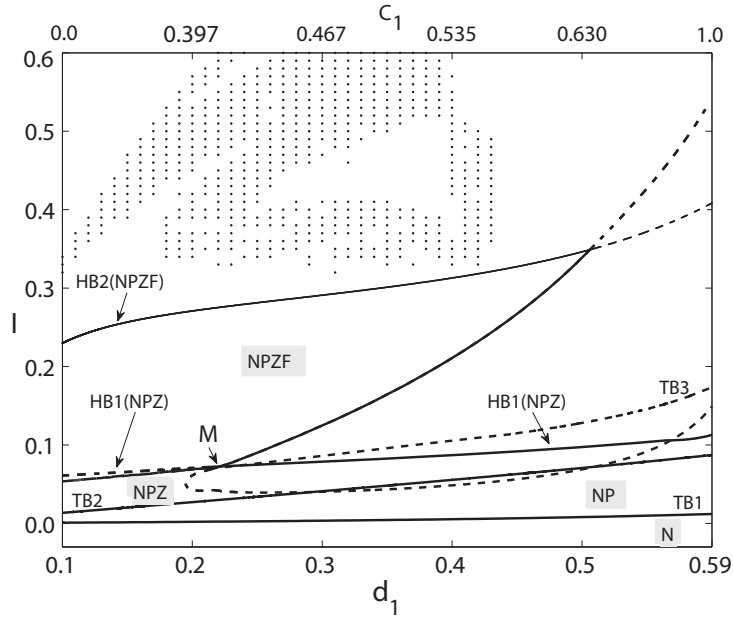} \quad
       \caption{Some of the bifurcations occurring as a function
of nutrient input $I$ and the phytoplankton mortality $d_1$, in
the range of values determined by the presence of contaminant
$C_1$ affecting phytoplankton. Values of $C_1$ are also indicated
in the upper horizontal axis. The name of the bifurcation lines is
indicated (for the case of the Hopf lines HB1, HB2 and HB3, the
name of the fixed point involved in the bifurcation is shown in
parenthesis). Crossing the continuous lines involves a qualitative
change for the state attained by the system. Inside regions
surrounded by continuous lines, the name of the relevant stable
fixed point is shown inside grey squares. Crossing the
discontinuous bifurcation lines does not involve a change in the
stable state (because, e.g., they correspond to bifurcations of
already unstable states). Immediately above the HB2 line, a limit
cycle involving all the species is the relevant attractor for low
values of $d_1$ (or $C_1$). The limit cycle on the $F=0$
hyperplane is the relevant attractor above the HB1 line for large
$d_1$. Additional bifurcations (not shown) occur in other regions
of the open areas above HB1 and HB2. M is a codimension-2 point
described in the main text. The dotted region identifies areas
where chaotic solutions have been found.
       } \label{fig4-xicont1}
\end{center}
\end{figure}

We now introduce contaminant $C_{1}$. It increases the mortality
of phytoplankton according to expression (\ref{6-dj}) for $i=1$.
The main bifurcations are shown in the 2-parameter bifurcation
diagram of Fig. \ref{fig4-xicont1} as a function of $d_1$ and $I$
(with values of $C_1$ also indicated). Expressions for the
bifurcation lines TB1, TB2 and TB3 as a function of $I$ and
$C_{1}$ can be obtained simply by replacing the mortality
(\ref{6-dj}) into the corresponding expressions (\ref{13-TB1}),
(\ref{14-TB2}), and (\ref{15-TB3}), respectively. The same can be
done numerically for the Hopf bifurcation lines HB1 and HB2.
Because of the sigmoidal effect of the contaminant (\ref{6-dj}),
its impact is mild for $C_{1}\ll0.5$, and it will saturate for
$C_{1}\gg1$. Thus, in both limits the bifurcation lines would
become parallel to the $C_1$ axis if plotted in terms of $C_{1}$
and $I$. The interesting behaviour is for intermediate values of
$C_{1}$, to which most of Fig. \ref{fig4-xicont1} pertains. In
this range the bifurcation lines are displaced towards higher
values of $I$. That is, the first effect of the contaminant is to
stabilize the simplest solutions, the ones which are stable at
lower nutrient load, delaying until higher nutrient loads the
transitions to the most complex solutions.

But this stabilizing effect is different for the different
solutions, and the most important qualitative change occurs at
point M in Fig.\ref{fig4-xicont1}. It is a codimension-2 point at which the
transcritical bifurcation TB3, involving the NPZ and
the NPZF fixed points, and the Hopf bifurcation HB1 of the NPZ point,
meet. A new Hopf bifurcation line of the NPZF equilibrium, HB3,
emerges also from that point. The cycle created at HB3 consists in oscillations
of all the four variables, similar to the cycle created at HB2.
Other characteristics of the organizing center M is that the Hopf bifurcations
change from subcritical to supercritical character across it, and also that a
line (not shown) of saddle-node bifurcations of the cycles created at HB1
and HB3 emerges also from M. There are a number of additional structures
in parameter space emerging from double-Hopf points, and transcritical
bifurcations of cycles which we do not describe further here.

\begin{figure}[p]
\begin{center}
    \includegraphics[width=.85\columnwidth,angle=0]{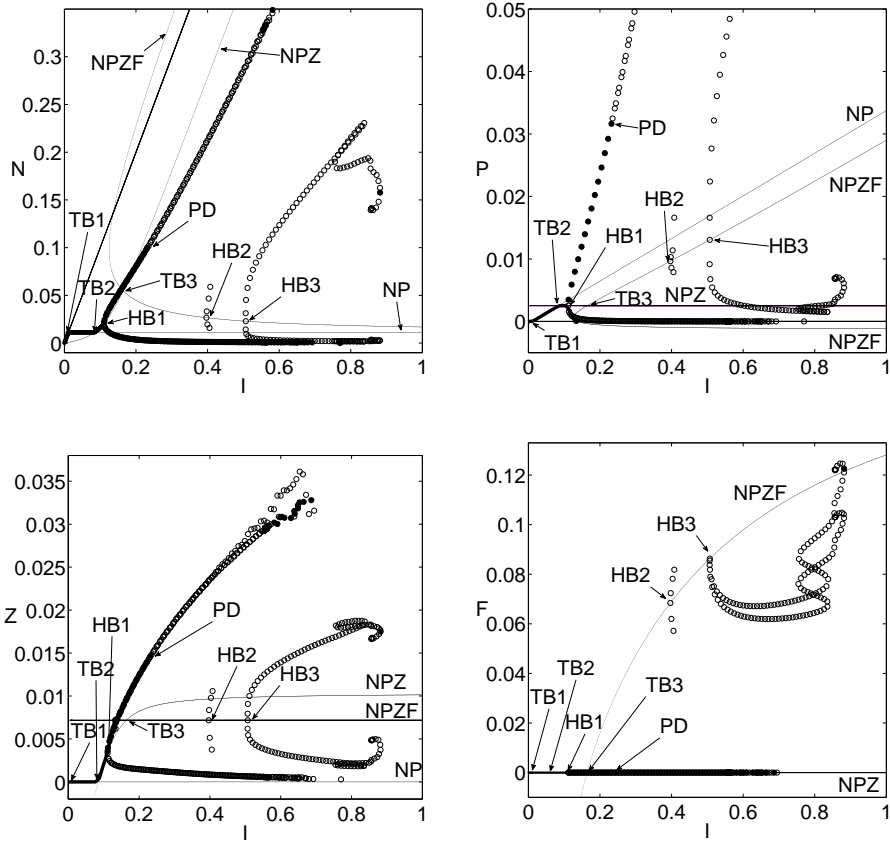}\\
   \caption{Bifurcation diagrams of the four variables as a
function of nutrient input parameter $I$, at a constant high value
of the contaminant affecting phytoplankton, $C_1=0.9$
($d_1=0.586$). Thick lines and full symbols denote stable fixed
points and maxima and minima of stable cycles, respectively, and
thin lines and open symbols, unstable ones. The name of the fixed
points is shown. The bifurcation points are identified by arrows.
PD is a period doubling bifurcation.}
   \label{fig5-cont1}
\end{center}
\end{figure}

Despite the complexity of the above scenario, its effect on the
bifurcation sequence when increasing nutrient level (up to
moderate levels) in the presence of contaminant values beyond M is
rather simple (see Fig.\ref{fig5-cont1}): since the lines TB3 and
HB1 have interchanged order, the Hopf bifurcation HB1 in which a
stable limit cycle is created in the hyperplane $F=0$ occurs
before the appearance of a positive NPZF equilibrium. As a
consequence, fish remains absent from the system even at
relatively high nutrient levels. This is one of the
counterintuitive outcomes of indirect effects:  adding a substance
which is toxic for phytoplankton makes fish disappear, whereas the
oscillating phytoplankton levels are indeed comparable with the
ones at zero contaminant (see Fig. \ref{fig5-cont1}). As in the
absence of a contaminant, period doublings and transition to chaos,
which we have not analyzed in detail, occur with further increases
in the value of $I$.

\begin{figure}[p]
\begin{center}
    \includegraphics[width=.85\columnwidth,angle=0]{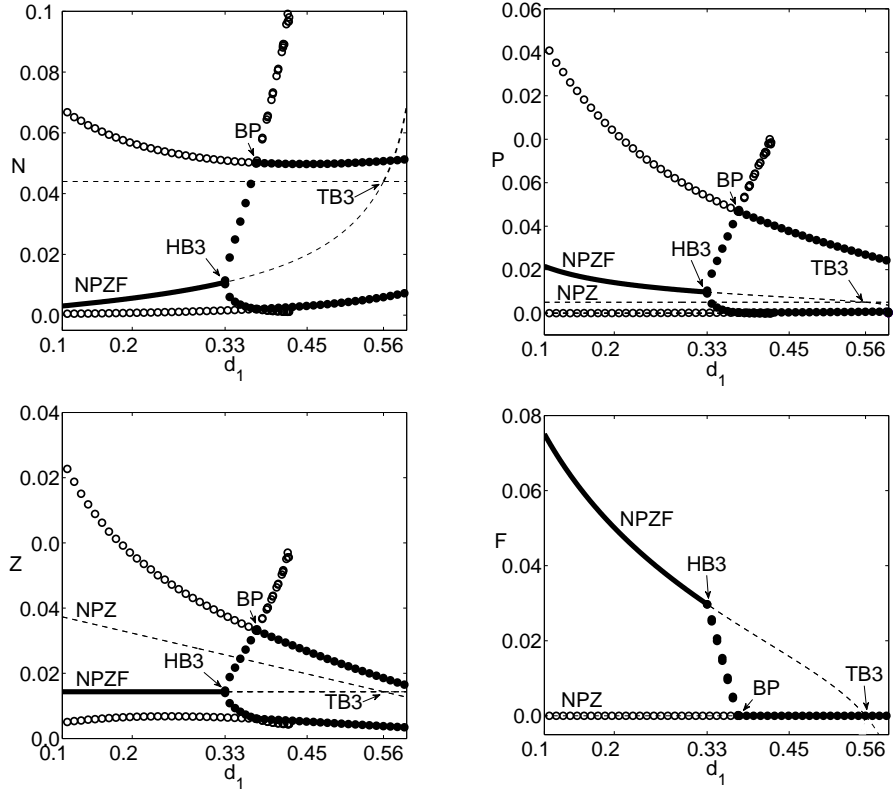}\\
   \caption{Bifurcation diagrams of the four variables as a
   function of $d_1$, affected by contaminant $C_1$, at constant
   nutrient input $I=0.15 ~mgN/l$. Thick lines and full
   symbols denote stable fixed points and maxima and minima of stable
   cycles, respectively, and thin lines
   and open symbols, unstable ones. BP is a transcritical
   bifurcation of cycles. The name of the fixed points is shown.
   The bifurcation points are identified by arrows.  }
 \label{fig6-cont1}
\end{center}
\end{figure}

A different view of the transitions can be given by describing the
bifurcations occurring by increasing the contaminant concentration
(or $d_1$) at constant $I$. Fig. \ref{fig6-cont1} shows that for
an intermediate value of the nutrient load, $I=0.15~mgN/l$. The
NPZF fixed point is stable at low contaminant, but oscillations
appear when crossing the HB3 lines. Very shortly after that, this
limit cycle involving all species approaches the $F=0$ hyperplane
until colliding with the cycle lying on that plane, which involves
only the $N$, $P$, and $Z$ species. At this transcritical
bifurcation, this limit cycle from which fish is absent becomes
stable and is the observed solution for larger $C_1$ or $d_1$. As
before, adding a substance which is toxic for the bottom of the
trophic chain has the main effect of suppressing the top-predator.

\subsection{Contaminant toxic to zooplankton}

\begin{figure}[p]
\begin{center}
   \includegraphics[width=0.8\columnwidth,angle=0]{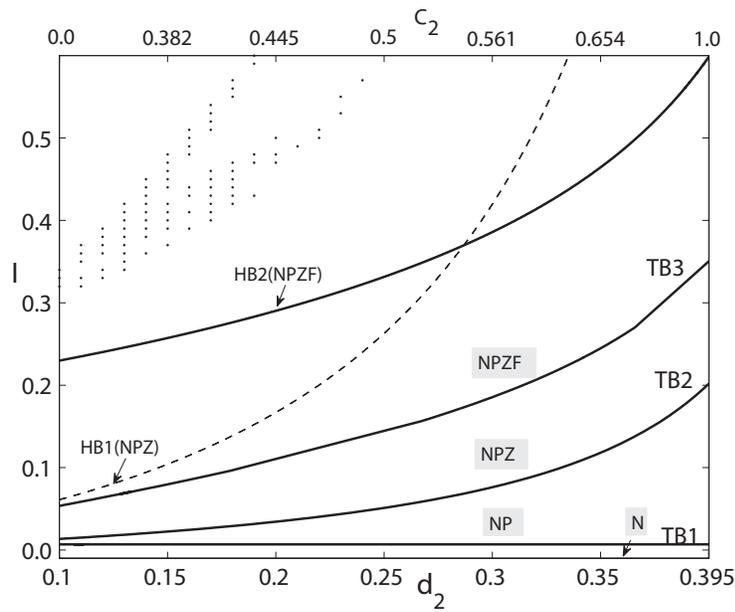} \quad
       \caption{Some of the bifurcations occurring as a function
of nutrient input $I$ and zooplankton mortality $d_2$, in the
range of values determined by the presence of contaminant $C_2$
affecting zooplankton. Values of $C_2$ are also indicated in the
upper horizontal axis. Names of fixed points and bifurcation lines
as in Fig. 4, as well as the meaning of continuous and
discontinuous lines. Immediately above the HB2 line, the relevant
attractor is a limit cycle involving all the species. Additional
bifurcations (not shown) occur at higher values of $I$. The dotted
region identifies areas where chaotic solutions have been found.}
\label{fig7-xicont2}
\end{center}
\end{figure}

The 2-parameter bifurcation diagram of Fig. \ref{fig7-xicont2}
displays the behaviour as a function of $I$ and the zooplankton
mortality $d_2$, as affected by contaminant $C_2$. As before, the
mortality expression (\ref{6-dj}) for $j=2$ can be inserted in the
expressions (analytical or numerical) for the bifurcations TB1,
TB2, TB3, HB1, and HB2 to elucidate the impact of the contaminant
$C_{2}$, acting on zooplankton, on the food chain dynamics. As
in the $C_{1}$ case, the bifurcation lines become displaced to
higher nutrient load values, so that the sequence of bifurcations
of Fig. \ref{fig3-nutrientes} becomes delayed to higher values of
$I$. In contrast with the $C_1$ case, the TB3 and HB1 lines do
not cross, so that there are no further qualitative changes with
respect to the case without contaminants (Fig.
\ref{fig3-nutrientes}), at least up to moderate values of $I$.

\begin{figure}[p]
\begin{center}
   \includegraphics[width=.85\columnwidth,angle=0]{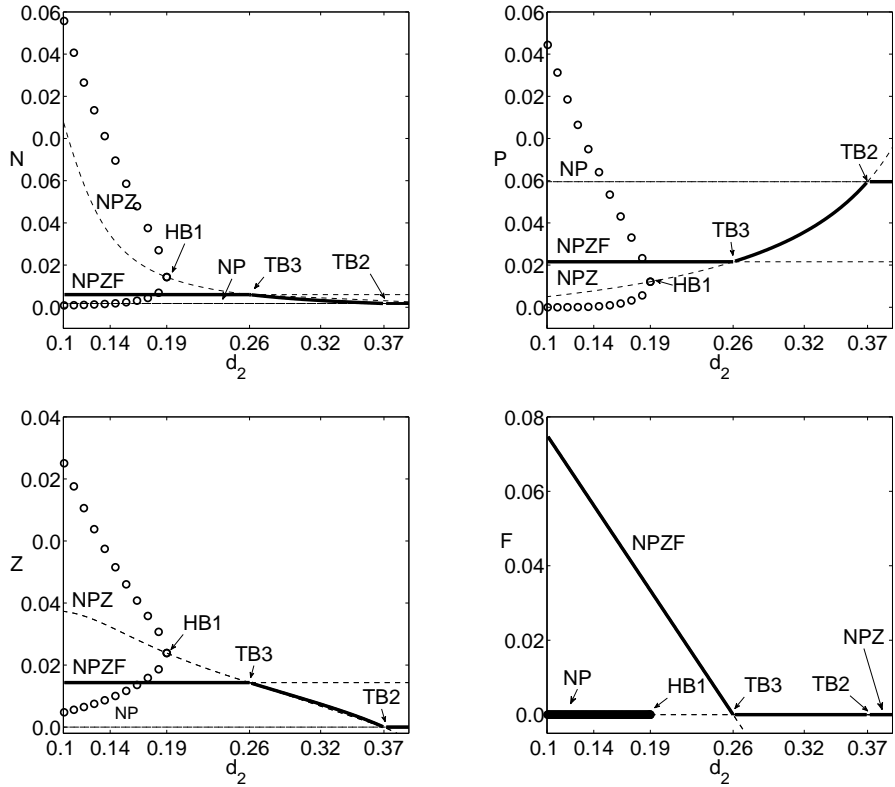}\\
   \caption{Bifurcation diagrams of the four variables as a function of $d_2$,
   affected by contaminant $C_2$, for a constant nutrient load $I=0.15 ~mgN/l$. Thick lines denote stable fixed and
   thin lines and open symbols, unstable fixed points and maxima and minima of
   unstable cycles. The name of the fixed points is shown. The bifurcation points
   are identified by arrows.} \label{fig8-cont2}
\end{center}
\end{figure}

Another view of the consequences of Fig. \ref{fig7-xicont2} can be
seen in Fig. \ref{fig8-cont2}, which shows the different regimes
attained at a fixed intermediate value of $I$ ($I=0.15~mgN/l$) and
increasing $C_{2}$ or $d_2$. The most remarkable indirect effect
is that, for $d_{2}<d_{2}^{TB3}=0.2592 ~ day^{-1}$
($C_{2}<C_{2}^{TB3}=0.5103$), zooplankton remains constant despite
the increase of $C_2$ which is toxic to it. The net effect of
$C_2$ in this range is to decrease the amount of fish until
extinction. Only for $C_{2}>C_{2}^{TB3}$ contaminant kills
zooplankton until extinction at $d_{2}=d_{2}^{TB2}=0.374 ~
day^{-1}$ ($C_{2}=C_{2}^{TB2}=0.7406$).

\subsection{Contaminant toxic to fish}

\begin{figure}[p]
\begin{center}
    \includegraphics[width=0.8\columnwidth,angle=0]{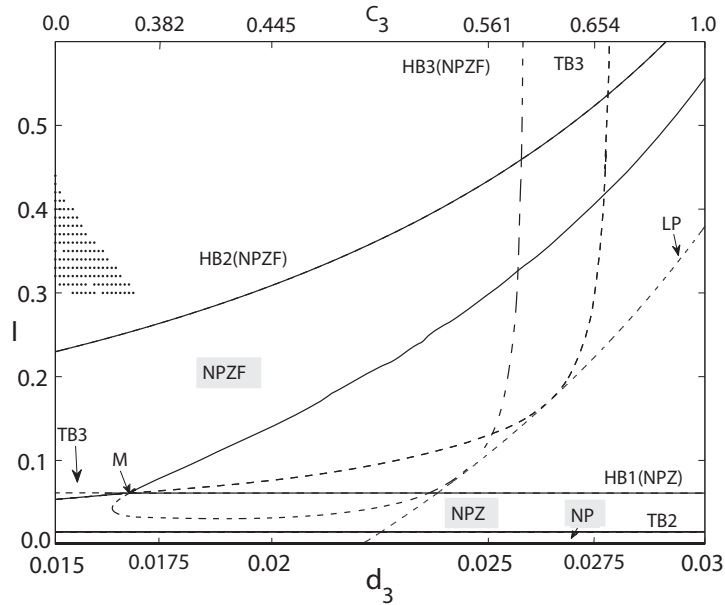} \quad
       \caption{Some of the bifurcations occurring as a function
of nutrient input $I$ and fish mortality $d_3$, in the range of
values determined by the presence of contaminant $C_3$ affecting
fish. Values of $C_3$ are also indicated in the upper horizontal
axis. Names of fixed points and bifurcation lines as in Fig. 4, as
well as the meaning of continuous and discontinuous lines.
Immediately above the HB2 line, the relevant attractor is a limit
cycle involving all the species. Additional bifurcations (not
shown) occur at higher values of $I$. The dotted region identifies
areas where chaotic solutions have been found. There is a region
of the area labelled as NPZF in which this stable fixed point
coexists with a stable limit cycle on the $F=0$ hyperplane.
       } \label{fig9-xicont3}
\end{center}
\end{figure}

\begin{figure}[p]
\begin{center}
      \includegraphics[width=.85\columnwidth,angle=0]{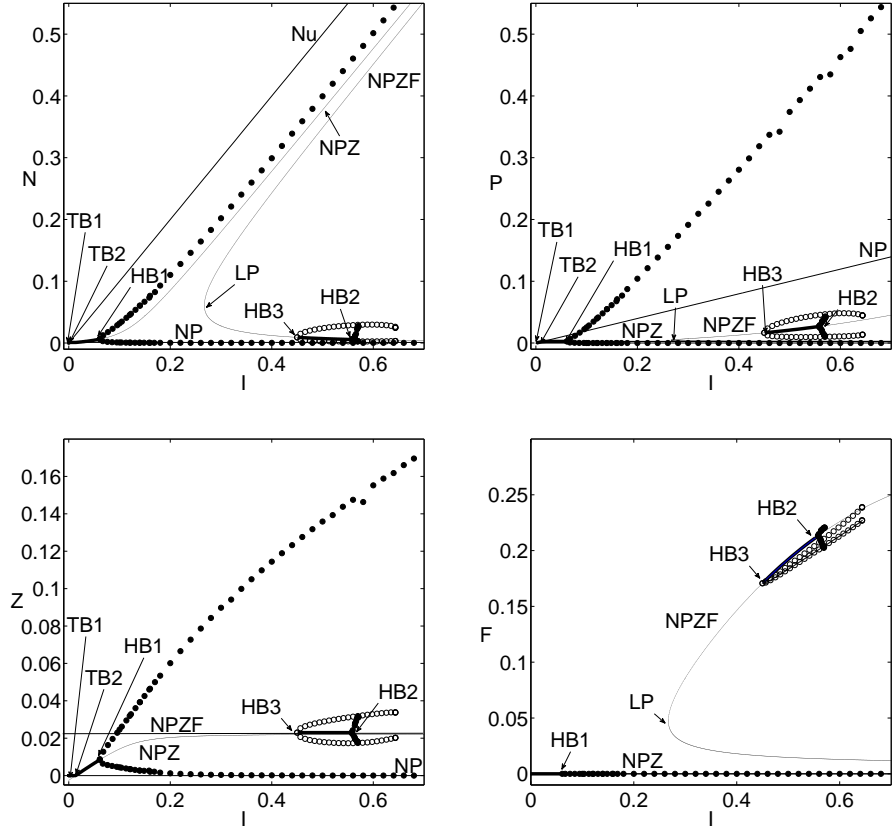}\\
   \caption{Bifurcation diagrams of the four variables as a
   function of nutrient input rate parameter $I$ for a high value
   of the contaminant affecting fish, $C_3=0.7$. Thick lines and
   full symbols denote stable fixed points and maxima and minima
   of stable cycles, respectively, and thin lines and open
   symbols, unstable ones. The name of the fixed points is shown.
   The bifurcation points are identified by arrows. There is a
   small region of coexistence (between HB3 and HB2) of the stable
   NPZF fixed point and a stable limit cycle on the $F=0$
   hyperplane, which leads later to coexistence of two limit
   cycles.} \label{fig10-cont3}
\end{center}
\end{figure}

The simplest bifurcation lines are shown in Fig.
\ref{fig9-xicont3} as a function of $I$ and $d_{3}$, the fish
mortality affected by contaminant $C_3$. As in the cases before,
bifurcations are delayed to higher values of $I$ when contaminant
is present. As in the $C_{1}$ case, this delay is different for
the different lines, resulting in a crossing of TB3 and HB1 in a
codimension-2 point M, joining there also to a new Hopf bifurcation
HB3 of the NPZF fixed point and other bifurcation lines (not
shown). Additional structures emerging from other codimension-2
points, such as double-Hopf points are also present but we do not
study them in detail. The qualitative behaviour when increasing
$I$ at large $C_{3}$ or $d_3$ (Fig. \ref{fig10-cont3}) is similar
to the $C_{1}$ case: there is a succession of N, NP and NPZ fixed
points followed by a Hopf bifurcation which leads to oscillations
of the $N$, $P$ and $Z$ variables, maintaining the absence of fish from
the system. Only at relatively high nutrient values does the NPZF
steady state become stable at the subcritical branch of the Hopf
bifurcation HB3 before becoming unstable again at HB2. Two of the
NPZF internal solutions (\ref{12-NPZF}), which, in contrast with
the $C_{3}=0$ case, are positive here, collide at a limit point.
In Fig. \ref{fig9-xicont3} the line of these points as a function
of the $I$ and $d_{3}$ parameters is labelled as LP. The two
solutions exist above that line, and cease to exist below. The
sequence of bifurcations encountered when increasing $C_{3}$ or
$d_3$ at constant intermediate values of $I$ is also similar to
the $C_{1}$ case of Fig. \ref{fig6-cont1} in that the NPZF stable
fixed point becomes a cycle involving all the variables when HB3
is crossed, and in that it approaches the $F=0$ plane shortly
afterwards. The details are, however, more complex because of the
presence in phase space of additional unstable cycles.

\section{Discussion and conclusion}

Because of the assumed sigmoidal influence of contaminant on
mortality, toxic effects on our food chain model have a distinct
effect at low and at large concentrations, with rather fast
transition behaviour in between.

At small and moderate contaminant concentrations the main effect
is a displacement of the different bifurcations towards higher
nutrient load values, so that transitions to states containing
less active compartments, and states without oscillations, become
relatively stabilized. Contaminants increase the stability of the
food chain with respect to oscillations caused by increased
nutrient input. A similar outcome has been observed in \cite{Upadhayay2005} for a
food-chain model composed of a toxin producing phytoplankton,
zooplankton and fish population. In that study chaotic dynamics
can be stabilized by increasing the strength of toxic substance in
the system.

For higher contaminant values, in most of the cases there is a
reordering of the different transitions, giving rise to the
appearance of new bifurcations which change qualitatively the
sequence of states encountered by increasing nutrient input. The
main effect, even in the cases in which such reordering does not
occur (the case of $C_2$ contaminant), is that the top predator
becomes the most depleted, being even brought to extinction. This
strong impact of the contaminant on the higher predator occurs
even in the cases in which the direct toxic effect is on lower
trophic levels. It seems that the increased mortality at lower
trophic levels becomes nearly compensated by a debilitation of
top-down control exerted by higher predators. Obviously, the top
predator can not benefit from this mechanism, thus becoming the
most affected.

Extrapolation of the above findings for real ecosystems may be
problematic, because of the much higher food web complexity found
in nature. We believe however that the counterintuitive indirect
effects described above should be kept in mind when analyzing the
complex responses that ecosystems display to changes in external
drivers such as nutrient load and pollutants, and that the
detailed identification of them performed here can help to
interpret some aspects of the behaviour of real ecosystems.

\section*{Acknowledgements}

We acknowledge support from the European Commission through the
integrated project THRESHOLDS (contract 003933) and from MEC
(Spain) and FEDER, through project FISICOS (FIS2007-60327).

\newpage




\newpage

\begin{table}[p]
    \caption{Parameters of the CC model.}
    \label{parameter-CC}
    \begin{center}
       \begin{tabular}{l l l l}
                  Parameters & & value& Units\\\hline
          Nutrient input  & $I$ & $0.15$ & mg N/l \\
         Inflow/outflow rate & $D$ & $0.02$ & $day^{-1}$  \\
         Max predation rate & $a_{1}$ & $1.00$ & $day^{-1}$  \\
                            & $a_{2}$ & $0.50$ & $day^{-1}$ \\
                            & $a_{3}$ & $0.047$ & $day^{-1}$ \\
          Half saturation cont & $b_{1}$ & $0.008$ & mg N/l \\
                          & $b_{2}$ & $0.01$ & mg N/l \\
                          & $b_{3}$ & $0.015$ & mg N/l \\
    Efficiency & $e_{1}$ & $1.00$& - \\
                & $e_{2}$ & $1.00$& - \\
             & $e_{3}$ & $1.00$& - \\
    Mortality(base values) & $d_{1}$ & $0.10$ &$day^{-1}$ \\
            & $d_{2}$ & $0.10$ &$day^{-1}$ \\
            & $d_{3}$ & $0.015$ &$day^{-1}$ \\
  Flushing rate & $f_{1}$ & $0.01$ &$day^{-1}$\\
               & $f_{2}$ & $0.01$ &$day^{-1}$\\
               & $f_{3}$ & $0.01$ &$day^{-1}$\\
     \end{tabular}
    \end{center}
\end{table}

\begin{table}[p]
    \caption{Contaminant parameters for the three compartments,
    $j=1,2,3$.}\label{parameter-contaminant}
    \begin{tabular}{|l|c|c|}
      \hline
      $j$ & $d^{(0)}_{j}$ & $\Delta d _{j}$ \\
      \hline
      1 (prey) & 0.1 & 0.5 \\
      2 (predator) & 0.1& 0.3 \\
      3 (top-predator) & 0.015 & 0.015 \\
      \hline
    \end{tabular}
\end{table}

\clearpage   


\section*{Figure captions}

\textbf{Fig.1.} Sigmoidal response of mortality to the
concentration of the toxic contaminant, according to Eq. (6).

\textbf{Fig.2.} (a). Projection on the PZF subspace of a
trajectory which starts close to the NP fixed point, approaches
the NPZ one, and finally is attracted by the NPZF fixed point.
$I=0.4~mgN/l$, $C_1=C_3=0$, and $C_2=0.8$. This shows the
approximate location of these points and that only the NPZF one is
stable for these parameter values. (b) Cyclic behaviour: Thick
line is a trajectory leading to an attracting limit cycle on the
NPZ hyperplane for $I=0.1~mgN/l$, $C_1=C_2=0$, and $C_3=0.8$ ;
dotted line is a trajectory attracted by the limit cycle involving
all the variables for $I=0.24~mgN/l$, $C_1=C_2=0$, and $C_3 =0.2$.

\textbf{Fig.3.} Bifurcation diagrams of the four variables as a
function of nutrient input parameter $I$ in the absence of
contaminants. Thick lines and full symbols denote stable fixed
points and maxima and minima of stable cycles, respectively, and
thin lines and open symbols, unstable ones. The names of the fixed
points are indicated. The relevant bifurcations (described in the
main text) occur at $I^{TB1}=0.0008909 ~mgN/l$, $I^{TB2}=0.01345
~mgN/l$, $I^{TB3}= 0.05352 ~mgN/l$, $I^{HB1}= 0.06101 ~mgN/l$, and
$I^{HB2}=0.2298 ~mgN/l$, locations which are indicated by arrows.

\textbf{Fig.4.} Some of the bifurcations occurring as a function
of nutrient input $I$ and the phytoplankton mortality $d_1$, in
the range of values determined by the presence of contaminant
$C_1$ affecting phytoplankton. Values of $C_1$ are also indicated
in the upper horizontal axis. The names of the bifurcation lines are
indicated (for the case of the Hopf lines HB1, HB2 and HB3, the
name of the fixed point involved in the bifurcation is shown in
parenthesis). Crossing the continuous lines involves a qualitative
change for the state attained by the system. Inside regions
surrounded by continuous lines, the name of the relevant stable
fixed point is shown inside grey squares. Crossing the
discontinuous bifurcation lines does not involve a change in the
stable state (because, e.g., they correspond to bifurcations of
already unstable states). Immediately above the HB2 line, a limit
cycle involving all the species is the relevant attractor for low
values of $d_1$ (or $C_1$). The limit cycle on the $F=0$
hyperplane is the relevant attractor above the HB1 line for large
$d_1$. Additional bifurcations (not shown) occur in other regions
of the open areas above HB1 and HB2. M is a codimension-2 point
described in the main text. The dotted region identifies areas
where chaotic solutions have been found.

\textbf{Fig.5.} Bifurcation diagrams of the four variables as a
function of nutrient input parameter $I$, at a constant high value
of the contaminant affecting phytoplankton, $C_1=0.9$
($d_1=0.586$). Thick lines and full symbols denote stable fixed
points and maxima and minima of stable cycles, respectively, and
thin lines and open symbols, unstable ones. The names of the fixed
points are shown. The bifurcation points are identified by arrows.
PD is a period doubling bifurcation.

\textbf{Fig.6.} Bifurcation diagrams of the four variables as a
function of $d_1$, affected by contaminant $C_1$, at constant
nutrient input $I=0.15 ~mgN/l$. Thick lines and full symbols
denote stable fixed points and maxima and minima of stable cycles,
respectively, and thin lines and open symbols, unstable ones. BP
is a transcritical bifurcation of cycles. The name of the fixed
points is shown. The bifurcation points are identified by arrows.

\textbf{Fig. 7.} Some of the bifurcations occurring as a function
of nutrient input $I$ and zooplankton mortality $d_2$, in the
range of values determined by the presence of contaminant $C_2$
affecting zooplankton. Values of $C_2$ are also indicated in the
upper horizontal axis. Names of fixed points and bifurcation lines
are as in Fig. 4, as well as the meaning of continuous and
discontinuous lines. Immediately above the HB2 line, the relevant
attractor is a limit cycle involving all the species. Additional
bifurcations (not shown) occur at higher values of $I$. The dotted
region identifies areas where chaotic solutions have been found.

\textbf{Fig. 8.} Bifurcation diagrams of the four variables as a
function of $d_2$, affected by contaminant $C_2$, for a constant
nutrient load $I=0.15 ~mgN/l$. Thick lines denote stable fixed and
thin lines and open symbols, unstable fixed points and maxima and
minima of unstable cycles. The names of the fixed points are shown.
The bifurcation points are identified by arrows.

\textbf{Fig. 9.} Some of the bifurcations occurring as a function
of nutrient input $I$ and fish mortality $d_3$, in the range of
values determined by the presence of contaminant $C_3$ affecting
fish. Values of $C_3$ are also indicated in the upper horizontal
axis. Names of fixed points and bifurcation lines are as in Fig. 4, as
well as the meaning of continuous and discontinuous lines.
Immediately above the HB2 line, the relevant attractor is a limit
cycle involving all the species. Additional bifurcations (not
shown) occur at higher values of $I$. The dotted region identifies
areas where chaotic solutions have been found. There is a region
of the area labelled as NPZF in which this stable fixed point
coexists with a stable limit cycle on the $F=0$ hyperplane.

\textbf{Fig. 10.} Bifurcation diagrams of the four variables as a
function of nutrient input rate parameter $I$ for a high value of
the contaminant affecting fish, $C_3=0.7$. Thick lines and full
symbols denote stable fixed points and maxima and minima of stable
cycles, respectively, and thin lines and open symbols, unstable
ones. The names of the fixed points are shown. The bifurcation
points are identified by arrows. There is a small region of
coexistence (between HB3 and HB2) of the stable NPZF fixed point
and a stable limit cycle on the $F=0$ hyperplane, which leads
later to coexistence of two limit cycles.


\end{document}